\documentclass{jfm_modified_2021}
\usepackage{graphicx}
\usepackage{amsmath, amssymb}
\usepackage[inactive,tightpage]{preview}

\usepackage{microtype}
\usepackage[normalem]{ulem}
\usepackage{mathtools,bm}
\usepackage{cases}

\usepackage{esvect}

\usepackage{enumitem}

\pdfmapfile{+rsfso.map}
\DeclareSymbolFont{rsfso}{U}{rsfso}{m}{n}
\DeclareSymbolFontAlphabet{\mathscr}{rsfso}

\usepackage{pdflscape}
\usepackage{afterpage}
\usepackage{flafter}
\usepackage{rotating}
\usepackage{fancyhdr}
\fancypagestyle{lscape}{%
\fancyhf{} 
\fancyfoot[LE]{}
\fancyfoot[LO] {}
 
}

\usepackage{bm}
\usepackage{microtype}

\usepackage{hyperref}
\usepackage[usenames, dvipsnames]{xcolor}
\hypersetup{colorlinks=true,linkcolor=MidnightBlue,citecolor=MidnightBlue}
\usepackage{natbib}
\usepackage{array}
\usepackage{booktabs}
\usepackage{multirow}
\usepackage{siunitx}
\usepackage{mathtools}

\usepackage{cleveref}
\usepackage{caption}
\usepackage{subcaption}
\usepackage{tabularx}
\newcolumntype{Y}{>{\centering\arraybackslash}X}

\usepackage{tikz}
\usepackage{pgfplots}
\pgfplotsset{compat=newest}
\usepgfplotslibrary{fillbetween}
\usepackage[outline]{contour}
\usepackage{import}

\usepackage{lscape}

\newcommand{\dt}[1]{\frac{\mathrm{d} #1}{\mathrm{d} t}}
\newcommand{\dx}[1]{\frac{\mathrm{d} #1}{\mathrm{d} x}}
\newcommand{\dd}[2]{\frac{\mathrm{d} #1}{\mathrm{d} #2}}

\newcommand{\pt}[1]{\frac{\partial #1}{\partial t}}
\newcommand{\px}[1]{\frac{\partial #1}{\partial x}}
\newcommand{\ppx}[1]{\frac{\partial^2 #1}{\partial x^2}}

\newcommand{\up}[0]{\mathrm{up}}

\shorttitle{Asymptotic analysis of grid models}
\shortauthor{Morawiecki and Trinh}

\title{On the evaluation of grid and grid-to-grid rainfall-runoff models and their differences with physical benchmarks}

\author{Piotr Morawiecki\corresp{\email{pwm27@bath.ac.uk}}
 \and Philippe Trinh\corresp{\email{hppt20@bath.ac.uk}}}

\affiliation{
    Department of Mathematical Sciences, University of Bath, Bath BA2 7AY, UK
}

\pubyear{XXXX}
\volume{YYY}
\pagerange{xxx--xxx}
\date{\today~[Draft]}

\newcommand{\Pe}[0]{\mathrm{Pe}}
\newcommand{\tsat}[0]{t_\mathrm{sat}}

\begin{document}

\maketitle

\begin{abstract}
In the first part of our study, we demonstrated how a simple physical benchmark model can be used to assess assumptions of the conceptual models, based on a lumped Probability Distributed Model (PDM) formulated by~\cite{lamb1999calibration}. In this second part, we extend the scope of our study to distributed models, which aim to represent the spatial variability of model's elements (\emph{e.g.} input precipitation, soil moisture levels, flow components etc.). For demonstration purposes, we assess the assumptions of the Grid and Grid-to-Grid models, commonly used for flood real-time forecasting in the UK.
While the distributed character of these models is conceptually closer to the physical model, we demonstrate that its exact implementation leads to many qualitative and quantitative differences in the model behaviour. For example, we show that the main assumption, namely that the speed of surface and subsurface flow is constant, causes the Grid-to-Grid model to significantly misrepresent scenarios with no rainfall, leading to too fast river flow decay, and scenarios with upstream rainfall, failing to capture characteristic flash flood formation.
We argue that this analytical approach of finding fundamental differences between models may help us to develop more theoretically-justified rainfall-runoff models, \emph{e.g.} models that can better handle the two aforementioned scenarios and other scenarios in which the spatial dependence is crucial to properly represent the catchment dynamics.
\end{abstract}


\section{Introduction}

In the first part of this paper, we studied the Probability Distributed Model (PDM) proposed by \cite{lamb1999calibration}. The novelty of our approach was to compare this conceptual model to a simple physically-based benchmark that we developed in our earlier work \citep{paper1,paper2,paper3}. The main difference is that, unlike the spatially-distributed physical benchmark model, the PDM represents the entire catchment with a single soil moisture, surface (fast), and subsurface (slow) storages. Despite this difference in fundamental assumptions, the model allowed us to successfully reconstruct key features of the physical model, mostly properly explaining the underlying physical mechanism, with only a few identified differences.

The goal of this second part of the paper is to present a similar analysis applied to a distributed rainfall-runoff model, \emph{i.e.} a model that represents the spatial variation of model variables (such as rainfall, groundwater, and overland flow). The main motivation for using distributed rainfall-runoff models in hydrology is that they can be used with spatially-distributed input data, including precipitation data and the topography of the terrain. This way, depending on the spatial distribution of the precipitation, we can have a different time until reaching the peak outlet flow. This feature is not captured by the lumped PDM model discussed previously.

For demonstration purposes, we decided to focus on analysing the Grid Model by \cite{bell1998grid} and its successor, the Grid-to-Grid (G2G) Model by \cite{bell2007development}. They were chosen because of their popularity in the hydrologic community. These models are commonly used in the UK, for example, for operational flood forecasting services for England, Wales, and Scotland \citep{cole2010grid, price2012operational}, by the Australian Bureau of Meteorology \citep{acharya2019evaluation, wells2019distributed}, the Southeast Asian Weather and Climate Science for Service Partnership, and in a wide range of research projects, \emph{e.g.} by \cite{bell2009use}, \cite{bell2018marius}, \cite{formetta2018estimating}, and \cite{kay2021climate}.

The Grid and G2G models represent surface and subsurface flow as spatially-distributed storages, between which the water flows at a constant speed, which value is fitted to match the available data. This is a fundamentally different approach from the one used in the physical models. In those models the surface flow speed depends on the terrain properties and surface water height (as given by, \emph{e.g.} Manning's law), while the groundwater flow speed depends on its depth and soil/rock properties (as given by the Boussinesq equation). As we shall demonstrate in this work, this difference in fundamental modelling assumptions produces both qualitatively and quantitatively different model responses to rainfall events. Therefore, these conclusions apply not only to the Grid and G2G Models but also to a wide range of distributed conceptual models in which surface and subsurface flow are described with a kinematic wave approximation of constant speed.

The paper is structured as follows. In \cref{sec:formulation}, we introduce the Grid and Grid-to-Grid models based on the formulation in the publications in which they were originally proposed. In \cref{sec:assumption_comparison}, we highlight the similarities and differences between the assumptions of the grid models and the physical benchmark. The consequences of these differences are demonstrated and discussed in \cref{sec:consequences}. In \cref{sec:conclusions}, we summarise this study and discuss potential implications for hydrologic research.

The implementation of the Grid-to-Grid and physical benchmark models, as well as the scripts used to generate figures from this paper, are available in our GitHub repository \citep{github_rr}.

\section{Formulation of Grid and Grid-to-Grid models}
\label{sec:formulation}

The discrete version of the Grid-to-Grid model from the original paper is presented in \cref{app:discrete_grid_to_grid}. Here, we present its continuous form obtained in the limit $\Delta x\to 0$ and $\Delta t\to 0$. We write down the equations to represent our simplified benchmark model \citep{paper3}, in which we model surface and subsurface flow along a one-dimensional hillslope of width $L_x$ (see Figure 1 from the first part of this paper). All hillslope properties, such as its gradient, soil depth, saturated hydraulic conductivity, and surface roughness, remain constant.

\begin{figure}
    \centering
    \includegraphics{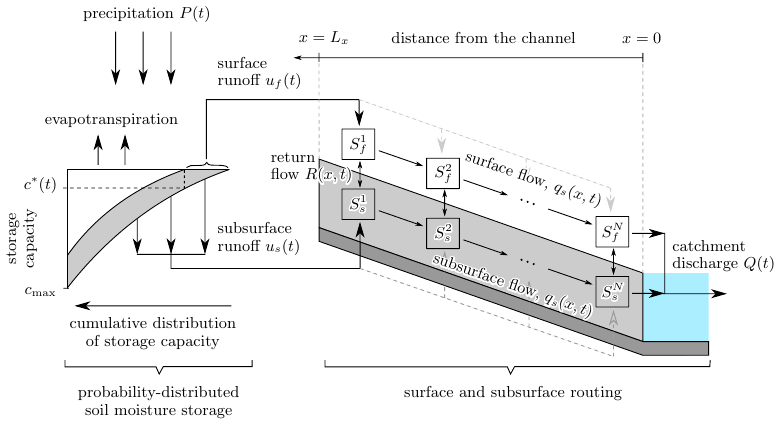}
    \caption{The structure of the Grid-to-Grid model by \cite{bell2007development} applied to 1D hillslope geometry used in our physical benchmark scenario. Probability-distributed soil moisture storage supplies surface and subsurface runoff to all stores distributed along the hillslope. In the case of discrete implementation, storages $S_f^i$ and $S_s^i$ for each grid element are used. In the continuous version, we represent surface and subsurface flow as a continuous function of position, $q_f(x,t)$ and $q_s(x,t)$.}
\end{figure}

The Grid-to-Grid model consists of three main components: soil moisture stores, surface stores, and subsurface stores, which can have different values in each grid element. In the continuous limit, we can represent them as a function of both time $t$ and location $x$. Like in the standard probability-distributed model (PDM), the soil moisture storage absorbs precipitation $P(x,t)$ and releases drainage $u_s(x,t)$. Its volume growth $c^*(x,t)$ is given by:
\begin{equation}
  \label{eq:c_ODE}
  \pt{c^*(x,t)}=\begin{cases}
  0 & \quad\text{for }c^*(x,t)=c_{\max}\text{ and }P(x,t)>u_s(x,t),\\
  P(x,t)-u_s(x,t) & \quad\text{otherwise},\\
  \end{cases}
\end{equation}
\emph{i.e.} it cannot exceed the $c_{\max}$ value. As in the case of our physical benchmark, we do not include the effect of evapotranspiration.

The drainage is dependent on the total soil moisture $S(c)$ as
\begin{equation}
    u_s(x,t)=\frac{S(c^*(x,t))^\beta}{k_g},
\end{equation}
where $k_g$ and $\beta$ are constant model parameters. Additionally, the surface runoff $r(t)$ is produced by all water that is not absorbed by the soil and is transferred to the groundwater store. From the water balance, we have
\begin{equation}
    u_f(x,t)=P(x,t)-u_s(x,t)-c_\text{max} \left(1-\frac{c^*(x,t)}{c_\text{max}}\right)^b\dt{c^*(x,t)}.
\end{equation}

The local drainage $u_s$ and surface runoff $u_f$ contribute to the formation of subsurface and surface flows, respectively. The pathway the flow follows is discussed in detail in the original papers, but in our simple catchment geometry, both flow components travel from $x=L_x$ towards $x=0$, where the channel is located. In this work, we are interested in the inflow to the river from a single hillslope, and the resulting channel flow is not considered. The flow transfer is described with a kinematic wave approximation assuming a constant surface water speed $c_f$ and a constant subsurface water speed $c_s$:
\begin{subequations}
    \label{eq:qs_qf_PDEs}
    \begin{align}
        \pt{q_s(x,t)}-c_s\px{q_s(x,t)}&=c_s\Big(u_s(t)-R(x,t)\Big), \\
        \pt{q_f(x,t)}-c_f\px{q_f(x,t)}&=c_f\Big(u_f(t)+R(x,t)\Big),
    \end{align}
\end{subequations}
with boundary conditions $q_s(L_x,t)=0$ and $q_f(L_x,t)=0$. The negative sign of the second (convection) term on the left-hand side represents the fact that the flow goes in the opposite direction to the x-axis (from $x=L_x$ to $x=0$). Apart from the surface runoff and drainage term, the source term includes a return flow, which is given by a fixed fraction of the subsurface flow: $R(x,t)=\frac{\gamma}{c_s} q_s(x,t)$, where $\gamma$ is a constant parameter (we used $\gamma$ instead of $r$ from the original paper to distinguish it from the precipitation rate denoted here by $r$).

To summarise, the Grid-to-Grid model involves seven parameters: $c_{\max}$, $b$, $k_g$, $\beta$, $c_s$, $c_f$, and $\gamma$, which are fitted to the available training data. The additional two parameters, $P_0$ and $P$, characterise our particular benchmark scenario. In the case of our benchmark scenario, the model formulation for the Grid Model is equivalent to the one presented here for no return flow ($\gamma=0$). Note that the return flow does not appear in our simplified physical benchmark model (\emph{i.e.} there is no exchange of flow between surface and subsurface flow), but it can in more complex geometries.

\section{Comparison of physical and grid models assumptions}
\label{sec:assumption_comparison}

In this section, we discuss the assumptions of the grid models (Grid and Grid-to-Grid model) by comparing them to the physical benchmark model. We investigate the assumptions of the runoff-production model and the routing model separately.

\subsection{Assumptions of the runoff-production model}
\label{sec:runoff_production_assumptions}

\begin{figure}
    \centering
    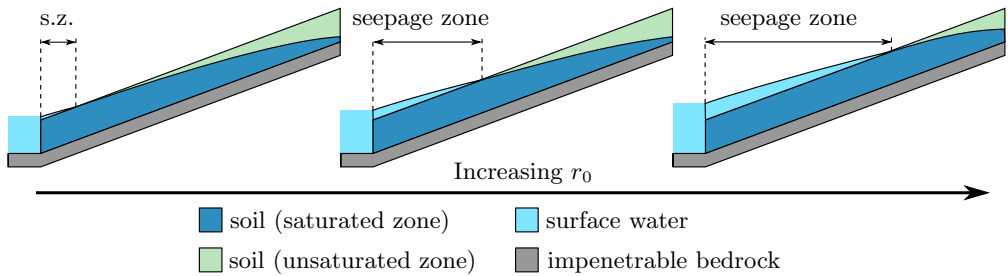
    \caption{Illustration showing the dependence between the mean rainfall rate $r_0$ and the size of the seepage zone. The dimensions of the hillslope and surface water height are not to scale.}
    \label{fig:increasing_saturated_zone}
\end{figure}

As we summarised in Section 2 of the first part of this paper, a key feature observed in our benchmark physical model is the existence of the saturation zone near the channel, which is the area of the catchment where groundwater reaches the surface. The size of this zone depends on the mean precipitation in the given season (see \cref{fig:increasing_saturated_zone}). Rainfall over this area accumulates as overland flow, which quickly travels to the channel. In the remaining part of the catchment, rainfall drains through the surface, reaching the groundwater table. The rise of the groundwater table can extend the size of the seepage zone, but this process is much slower.

The probability-distributed soil moisture (PDM) approach, as we demonstrated in the previous part, also predicts that part of the catchment is fully saturated. However, despite using a similar runoff-production model in grid models, it does not properly represent the spatial range of the seepage zone. In our benchmark scenario, each grid element's moisture storage is represented with the same PDM parameters, so for uniform rainfall, all grid elements produce both overland and groundwater flow at the same rate, regardless of their distance from the channel. This behaviour would be expected for Hortonian flow (\emph{i.e.} overland flow occurring when the precipitation rate exceeds the soil infiltration capacity), but not for the saturation excess overland flow that forms only in the seepage zone around the channel.

The saturation excess overland flow results from the interaction between surface and subsurface water, which is not included in the Grid Model. It is introduced in the Grid-to-Grid Model by adding a return flow from subsurface to surface stores. However, the flow generated in this way reaches the river with a much slower timescale, reflecting the variation of the groundwater flow. As a consequence, it does not lead to a rapid rise of overland flow during rainfall, as observed in our physical benchmark model. This effect can only be caused by direct surface runoff production, which is independent of the groundwater depth.

There is one significant practical consequence of this inconsistency in grid and physical model assumptions. According to the physical model, the longer the seepage zone is, the longer the early-time period during which the rainfall accumulated over the seepage zone reaches the river. This is why the critical time, given by
\begin{equation}
    \label{eq:tsat_dimensional}
    \tsat = \frac{L_z}{r}\left[\frac{S_x^{1/2}K_s n}{L_z^{k-1}} \left(\frac{L_x r}{K_s S_x L_z} - \frac{r}{r_0}\right)\right]^{1/k},
\end{equation}
increases with the mean rainfall value $r_0$. On the other hand, in the case of grid models, the time for water accumulated over the seepage zone to reach the channel is constant and equal to $t=L_x/c_f$. Therefore, we may expect the rate at which flow rises to be underestimated in dry seasons with low $r_0$ or overestimated in humid seasons with high $r_0$.

There are other fundamental differences between the PDM and physical benchmark models, such as the dependence on rainfall intensity $r$. Since these differences are a direct consequence of the PDM formulation, not the Grid-to-Grid spatially-distributed structure, they are discussed in detail in the first part of this paper.

\subsection{Comparison of kinematic wave equation and physical models}

Apart from the surface/subsurface production scheme, the equation used to describe their propagation is also different for grid and physical models. The kinematic wave equation, used in the Grid Model (and Grid-to-Grid Model for $\gamma=0$), to describe the surface and subsurface flow~\eqref{eq:qs_qf_PDEs}, has a general form:
\begin{equation}
    \label{eq:kinematic_wave_general}
    \pt{q(x,t)}+c\px{q(x,t)}=cu(x,t),
\end{equation}
where $q(x,t)$ represents the flow of surface or subsurface water, $c$ is its speed, and $u(t)$ is the source term (surface runoff or groundwater recharge). By expressing the flow $q$ as the product of surface/subsurface storage $s$ (water volume per unit area) and its propagation speed $c$, \emph{i.e.} $q=cs$, we obtain:
\begin{equation}
    \pt{s(x,t)}+\px{cs(x,t)}=u(t),
\end{equation}
which is a more standard form of the PDE for kinematic wave propagation. It resembles the formulation of governing equations commonly used in physical modelling, namely, the St. Venant equation for overland flow and the Boussinesq equation for groundwater flow, with a few key differences. In the St. Venant equations, the storage corresponds to the surface water height $s=h$. However, the speed of wave propagation is not constant, but is often assumed to depend on $h$. According to Manning's law, used in our Benchmark scenario, the relation is given by:
\begin{equation}
    \label{eq:Manning}
    c(h)=-\frac{\sqrt{S_x}}{n_s}h^{k-1},
\end{equation}
where $S_x$ is the elevation gradient along the hillslope, $n_s$ is Manning's constant characterising the surface roughness, and $k$ is an exponent typically equal to $k=\frac{5}{3}$. In the case of the Boussinesq equation on the hillslope, the speed is dependent on the gradient of $h$ as follows:
\begin{equation}
    c(h)=-K_s\left(S_x+\px{h}\right),
\end{equation}
where $K_s$ is the hydraulic conductivity of the saturated soil. Moreover, in this model, storage $s$ is not equivalent to $h$, but scales additionally with the drainable porosity $f$, defined as the mean fraction of the soil that can be filled with water, \emph{i.e.} $s=fh$.

To summarise, physical models are characterised by a nonlinear convection term. In contrast, the Grid and Grid-to-Grid Models assume a constant speed, which leads to a linear convection term. Losing the nonlinearity property may result in different hydrograph behaviour, the consequences of which are investigated in \cref{sec:consequences}.

\section{Consequences of different kinematic wave models}
\label{sec:consequences}

To highlight the key consequences, we discuss three benchmark scenarios. All of them are based on our 1D hillslope geometry but with different precipitation settings. In \cref{sec:scenario_uniform_rainfall}, we examine the hydrograph from the original scenario with uniform precipitation over the entire hillslope. In \cref{sec:scenario_no_rainfall}, we investigate the shape of the hydrograph representing a decrease in river inflow in the absence of rainfall. Finally, in \cref{sec:scenario_upstream_rainfall}, we explore the hydrograph obtained from rainfall over only the upstream part of the hillslope/catchment.

The goal of each scenario is to highlight different consequences of the linear kinematic wave approximation for surface flow. In order to focus only on the overland flow, we consider the flow only in the seepage zone. The late-time impact of seepage zone growth is not investigated.

\subsection{Impact of the simulated precipitation rate}
\label{sec:scenario_uniform_rainfall}

One of the main features of the physical model is the critical time over which the surface water, accumulating over the seepage zone, reaches the channel. Further growth in the inflow to the river is caused by the slowly rising groundwater. This timescale depends on the rainfall precipitation rate $r$, as given by equation~\eqref{eq:tsat_dimensional}, which scales as $\tsat\propto r^{1/k-1}=r^{-2/5}$.

\begin{figure}
    \centering
    \includegraphics{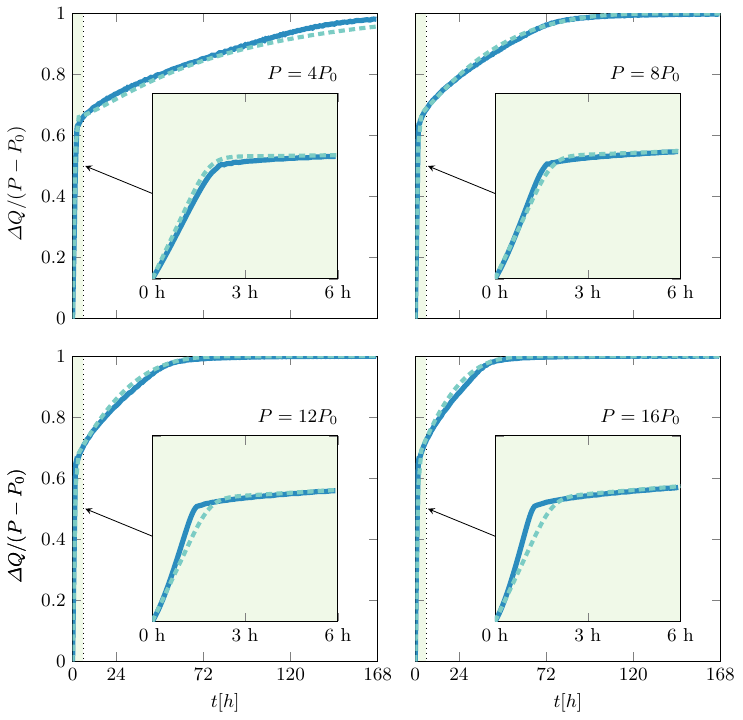}
    \caption{Hydrographs obtained for different rainfall rates $r$. Grid-to-Grid model parameters were fitted to minimise the mean square error averaged over all four presented hydrographs. The critical time for the Grid-to-Grid model was defined as the time at which the second derivative of $Q(t)$ is the highest.}
    \label{fig:r_dependence}
\end{figure}

This is because higher $r$ causes the height of the surface water to increase more, which, following Manning's law~\eqref{eq:Manning}, increases the overland flow speed. Consequently, the surface water reaches the river faster. The same argument applies to the channel flow. Therefore, for more extreme rainfalls, we should expect the flow at the catchment's outlet to rise much faster over a shorter timescale.

On the other hand, the speed of surface flow in the Grid-to-Grid Model is independent of the rainfall intensity and, consequently, the timescale remains constant. Therefore, similar to the lumped PDM from the previous section, we should expect that grid models fitted to standard rainfall events will underestimate the river flow growth rate during extreme rainfalls.

We investigated this effect by running four simulations of catchment response to a single rainfall event with different precipitation rates. The results are presented in \cref{fig:r_dependence}. Note how the timescale of the initial fast growth of river inflow decreases with $r$ according to the physical model but stays constant in the case of the Grid-to-Grid model. Nevertheless, the impact of this difference in timescales on the overall shape of the hydrograph is relatively small compared to the effects of using the Grid Model kinematic wave approximation in the benchmark scenarios presented in the next two sections.

\subsection{Impact on the asymptotic behaviour after the rainfall}
\label{sec:scenario_no_rainfall}

\begin{figure}
    \centering
    \includegraphics{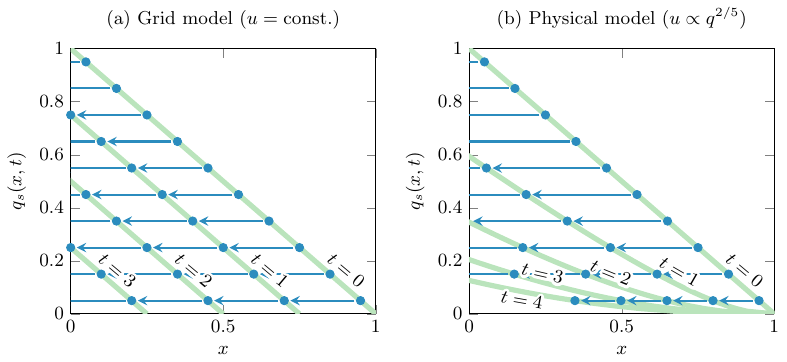}
    \caption{Illustration of overland profiles for grid models with a constant flow speed $u=0.5$ and for the Saint Venant equation with variable flow speed $u=q^{2/5}$. Note that in the first case, the river inflow reaches $0$ at $t=2$, while in the second case, the rate at which the flow decreases slows down.}
    \label{fig:drying_demo}
\end{figure}

The benchmark scenarios discussed so far have shown the impact of rainfall on the storm hydrograph. However, we also observe a significant difference between the hydrograph properties in the dry period after the rainfall. This is due to the fact that in grid models, the time required for the entire overland flow to reach the river is finite and given by $t_\mathrm{dry}=\frac{L_x}{c}$. According to Manning's equation, areas with low surface water height will take an arbitrarily long time to reach the stream (see \cref{fig:drying_demo}). Therefore, if there were no surface water infiltration into the soil, the overland flow would asymptotically approach zero as $t\rightarrow\infty$.

\begin{figure}
    \centering
    \includegraphics{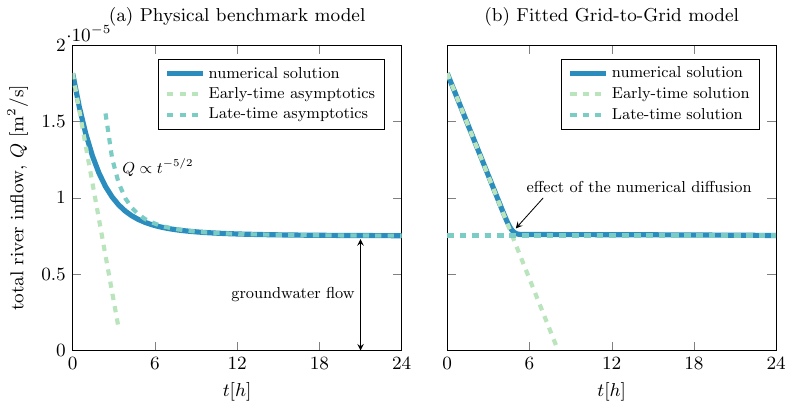}
    \caption{Overland component of the river inflow according to the physical benchmark and Grid-to-Grid models. The dashed lines represent different asymptotic approximations, as discussed in \cref{app:dry_scenario}.}
    \label{fig:dry_scenario}
\end{figure}

In \cref{app:dry_scenario}, we showed that for a hillslope of length $L_x$, grid models predict a linear decay of storm flow from $rL_x$ to $0$ at $t_\mathrm{dry}$, while according to the physical model, it initially decays linearly with time $t$, but the long-time asymptotic behaviour is $q\propto t^{-5/2}$ as $t\rightarrow\infty$. This behaviour can be observed in the numerical results presented in \cref{fig:dry_scenario}.

\subsection{Impact of the spatial distribution of the rainfall}
\label{sec:scenario_upstream_rainfall}

In the original three benchmark papers, only simulations for uniform rainfall were discussed. However, considering spatially distributed rainfall may introduce additional features to the hydrographs, and since Grid and Grid-to-Grid models were specifically designed to work with spatial precipitation data, we should discuss their consequences as well.

\begin{figure}
    \centering
    \includegraphics[width=\linewidth]{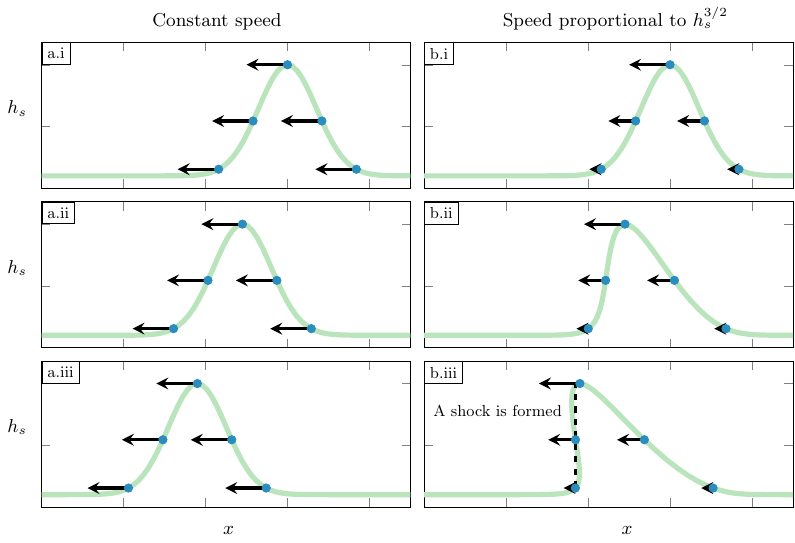}
    \caption{Illustration of shock formation using an example of a bell-shaped $h(x)$ function. In the figures on the left, the flow speed is constant (as in the Grid-to-Grid model), while in the figures on the right, the flow speed is proportional to $h_s^{2/3}$ (as in Manning's law used in the physical model). Five points on the profile were highlighted to show the movement of characteristic lines corresponding to these points in time. In the last plot, the red solid line represents the location of the shock being formed.}
    \label{fig:shock_formation}
\end{figure}

In the case of grid models, surface flow propagates at a constant speed, so any spatial surface flow profile maintains its shape (or grows if additionally fed by rainfall) as shown in \cref{fig:shock_formation}a. On the contrary, in the physical models, the points at which the surface height is propagating faster can cause the profile to stretch or compress as it propagates (see \cref{fig:shock_formation}b). This can lead to situations where high surface flow catches up with slow-moving flow. If not for the diffusive terms appearing in the diffusive approximation of the Saint Venant equations, this nonlinear behaviour leads to a shock formation (see \cref{fig:shock_formation}b.iii). This occurs in areas located downstream from the region covered by rainfall, occasionally leading to the creation of flash floods~\citep{hapuarachchi2011review}.

\begin{figure}
    \centering
    \includegraphics[width=\linewidth]{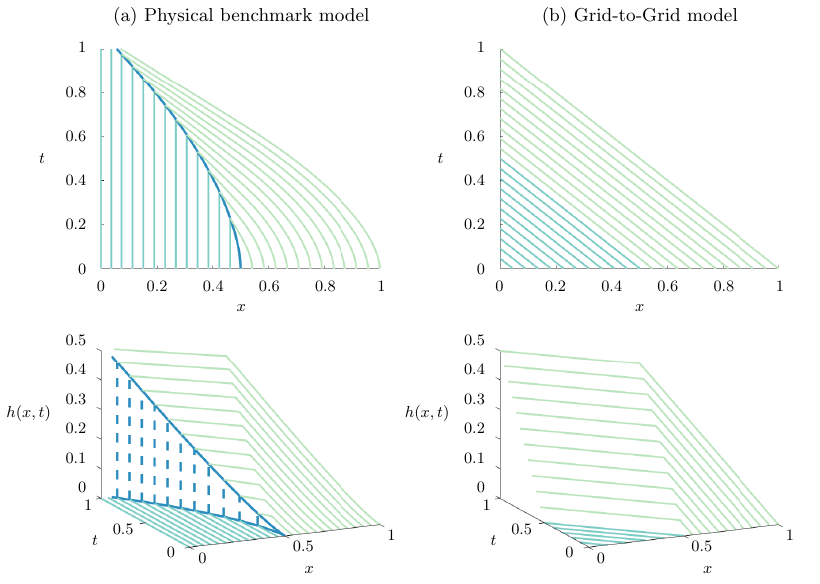}
    \caption{Characteristic diagram for the physical benchmark model and Grid-to-Grid model. Since cyan characteristic lines, which originate from the upstream region under rainfall $x\geq L_x/2$, intersect with yellow lines originating from the downstream region without rainfall $x<L_x/2$, a characteristic shock is formed. Such an intersection does not appear in the Grid-to-Grid model, since all characteristic lines are moving at an identical speed.}
    \label{fig:shock_diagram}
\end{figure}

This behaviour, which emerges naturally from physical models, cannot be observed in grid models. We can demonstrate this by considering a hillslope (or channel) in which we have uniform rainfall in the upper half and no rainfall in the lower half. We used the method of characteristics to find solutions for flow $q(x,t)$ for both the G2G Model and the physical benchmark model. Details of this analysis are presented in \cref{app:upstream_rainfall}. The solution is presented in \cref{fig:shock_diagram}. Note that the surface water height (or surface storage) at the river ($x=0$) rises gradually in the Grid-to-Grid model, while in the physical benchmark model, it rises instantly to a high value when the shock wave reaches the river.

\section{Conclusions}
\label{sec:conclusions}

\noindent In this work, we have demonstrated how simple physical benchmark models can be used to assess the underlying assumptions of conceptual rainfall-runoff models, specifically the Grid-to-Grid model commonly used in the UK. While the G2G model shares similarities with physical models in terms of separating precipitation into fast surface flow and slow subsurface flow, its simplistic structure leads to misrepresentation of some important features observed in the overland-dominated catchments.

We have identified two key assumptions that contribute to the observed differences. First, the runoff-production scheme in the G2G model is independent of the groundwater depth, causing it to overestimate overland flow when the groundwater level is high and underestimate it when the groundwater level is low. This limitation makes the G2G model vulnerable in regions with high seasonal variation in mean rainfall and evapotranspiration, as well as in regions experiencing long-term trends in groundwater levels, such as those affected by climate change.

The second assumption that introduces inconsistencies between models is related to the constant surface and subsurface speeds in the G2G model, while in physical models, these speeds depend on variables such as surface/subsurface water height and elevation gradient. This oversimplification has several consequences. First, the G2G model underestimates the rate of flow rise for extreme rainfalls, as the overland flow timescale does not depend on the precipitation rate. Second, the overland flow in the G2G model disappears after a finite time, while in physical models, a slow decay of overland flow is observed at late times, scaling as $t^{-5/2}$. Finally, the G2G model is not suitable for predicting flash floods, characterised by rapidly increasing flows when rainfall is limited to the upstream region of the catchment. Such dynamics can be explained by models with height-dependent surface flow speeds.

Although the benchmark scenario used in this study is a simplification of real-world systems, it allows us to identify key problems in the formulation of the G2G model and their downstream consequences. We argue that if essential flow features are misrepresented, even in the simplest catchment, we should expect that they may lead to inaccuracies when applied to real-world scenarios.

As \cite{kirchner2006getting} argued, "\textit{advancing the science of hydrology will require not only developing theories that get the right answers, but also testing whether they get the right answers for the right reasons}". Kirschner, as well as other authors (cf. for instance \cite{beven2018hypothesis}), has stressed the importance of examining model assumptions before relying upon model predictions. The approach presented in this work, combining asymptotic analysis with benchmarking, can help identify key fundamental differences between conceptual, physical, and statistical models of flooding, which is especially useful in conditions underrepresented in the available data. This methodology can hopefully yield a stronger foundation on which more theoretically-justified and reliable models can be developed.

\mbox{}\par
\noindent \textbf{Acknowledgements.} We thank Sean Longfield (Environmental Agency) for useful discussions, and for motivating this work via the 7th Integrative Think Tank hosted by the Statistical and Applied Mathematics CDT at Bath (SAMBa). We also thank Thomas Kjeldsen, Tristan Pryer, Keith Beven and Simon Dadson for insightful discussions. Piotr Morawiecki is supported by a scholarship from the EPSRC Centre for Doctoral Training in Statistical Applied Mathematics at Bath (SAMBa), under the project EP/S022945/1.

\bibliographystyle{plainnat}
\bibliography{bibliography}

\appendix

\section{Discrete version of Grid-to-Grid}
\label{app:discrete_grid_to_grid}

In this appendix, we present a discrete version of the Grid-to-Grid model by~\cite{bell2007development}, which we used to run numerical experiments. Let divide time into time steps of length $\Delta t$ and hillslope length into elements of length $\Delta x$. By index $i$, let us denote the value of each variable in the $i$\textsuperscript{th} time step. In each time step, we compute the soil moisture storage as
\begin{equation}
    c^{i+1}=\min\left(c_{\max},c^i+\Pi^i\Delta t\right)\quad \text{where }\Pi^i=P^i-d^i,
\end{equation}
and the resulting groundwater recharge $d^i$ and surface runoff $r^i$ following equation~(28) from~\cite{moore1985probability} as
\begin{equation}
    d^i=\frac{\left(S^i\right)^\beta}{k_g},\quad
    r^i = \Pi^i + \frac{S_{\max}}{\Delta t} \left[\left(1-\frac{c^{i+1}}{c_{\max}}\right)^{b+1}-\left(1-\frac{c^i}{c_{\max}}\right)^{b+1}\right].
\end{equation}
The fast and slow store values are updated using the following form of the discrete kinematic wave equations:
\begin{align}
    q_s^{i+1,j}&=(1-\theta_s)q_s^{i,j}+\theta_s\left(q_s^{i,j-1}+d^i-R^{i,j}\right), \\
    q_f^{i+1,j}&=(1-\theta_f)q_f^{i,j}+\theta_f\left(q_f^{i,j-1}+r^i+R^{i,j}\right).
\end{align}
Here, $q_s^{i,j}$ and $q_f^{i,j}$ denote the flow in the $j$\textsuperscript{th} grid element at time $i$, $\theta_s=\frac{c_s\Delta t}{\Delta x}$ and $\theta_f=\frac{c_f\Delta t}{\Delta x}$ are referred to as dimensionless wave speed (equivalent of the Courant number used in CFD), and $R^{i,j}=\frac{\gamma\Delta x}{c_s} q_s^{i,j}$ is a return flow. We take $q_s^{i,0}=q_f^{i,0}=0$ as a boundary condition at the catchment's border. The total inflow into the river (located at $j=N_x$) is given by the sum of the surface and subsurface flow components at the river:
\begin{equation}
    Q^i=q_s^{N_x,i}+q_f^{N_x,i}.
\end{equation}

In the limit $\Delta t\rightarrow 0$ and $\Delta x\rightarrow 0$, this model reduces to the continuous form presented in \cref{sec:formulation}.

\section{Solution for the upstream rainfall scenario}
\label{app:upstream_rainfall}

\noindent Let us consider a surface flow accumulating over a hillslope of length $L_x$, slope $S_x$, and surface runoff $U$ present only in the upstream part of the catchment, as given by
\begin{equation}
    u(x) = \begin{cases} 0 & x < x_\up \\ r & x \geq x_\up \end{cases}.
\end{equation}

In this appendix, we derive the solution of the kinematic wave approximation used in the Grid Model and Saint Venant equations for the overland flow. In order to focus only on the evaluation of the surface flow component, we will assume that there is no exchange between the surface and subsurface flow since, as we discussed in \cref{sec:runoff_production_assumptions}, there are differences in how such exchange is handled in grid and physical models. We will also assume that initially there is no surface water present over the entire hillslope.

\subsection{Grid models}

To solve the kinematic wave equation~\eqref{eq:kinematic_wave_general}, we use the method of characteristics, where the solution is given by characteristic curves $t(\tau)$, $x(\tau)$, and $q_s(\tau)$, with $\tau$ as a curve parameter. These characteristic curves represent the path that water follows, starting from any point on the hillslope at $t=0$ and from the catchment boundary $x=L_x$ at later times $t>0$.

These curves are governed by the Lagrange-Charpit equations:
\begin{equation}
    \dd{t}{\tau} = 1, \quad \dd{x}{\tau} = -c, \quad \dd{q}{\tau} = c u(x)
\end{equation}

The solution for curves starting at $x=x_0$ for $x_0\in\left[0,x_\up\right[$ is given by:
\begin{equation}
    t = \tau, \quad x = x_0-c\tau, \quad q = 0.
\end{equation}
The last of these lines reaches the river at time $t=\tau_1=\frac{x_\up}{c}$, and since the flow $q=0$ for all these curves, the river inflow will be zero for all $t\leq\frac{x_0}{c}$.

The solution for curves starting at $x=x_0$ for $x_0\in\left[x_\up,L_x\right]$ is given by:
\begin{equation}
    t = \tau, \quad x = x_0-c\tau, \quad q = c \int_0^\tau u(x(\xi)) \mathrm{d}\xi = \begin{cases} cr\tau & \tau < \tau_\up \\ cr\tau_\up & \tau \geq \tau_\up. \end{cases}
\end{equation}
Here, $\tau_\up=\frac{x_0-x_\up}{c}$ represents the time when the given characteristic curve passes through the point $x=x_\up$. The last of these lines reaches the river at time $t=\tau_2=\frac{L_x}{c}$. The resulting river inflow is given by:
\begin{equation}
    q(x=0,t) = r\left(ct-x_\up\right),
\end{equation}
so it linearly increases from $q(\tau_1)=0$ up to $q(\tau_2)=r\left(L_x-x_\up\right)$. For $t>\tau_2$, the characteristic curves will be the same as the last curve starting from $x_0=L_x$, but shifted in time. The resulting river inflow remains unchanged. In summary, the river inflow is given by:
\begin{equation}
    q(x=0,t) = \begin{cases} 0 & t \leq \frac{x_\up}{c} \\ r\left(ct-x_\up\right) & \frac{x_\up}{c} < t \leq \frac{L_x}{c} \\
    r\left(L_x-x_\up\right) & \frac{L_x}{c} < t
    \end{cases}
\end{equation}
The solution for the surface water height $h(x,t)=q(x,t)/c$ is presented in \cref{fig:shock_diagram}b.

\subsection{Physical benchmark}

In the case of our physical benchmark, the overland flow is governed by the Saint Venant equation:
\begin{equation}
    \pt{h}+\pt{c(h)h}=u(x).
\end{equation}
After substituting the Manning's law~\eqref{eq:Manning} for $c(h)$, we obtain:
\begin{equation}
    \label{eq:St_Venant}
    \pt{h}-kAh^{k-1}\pt{h}=u(x),
\end{equation}
where we introduced $A=\frac{\sqrt{S_x}}{n_s}$ as a constant for brevity. The solution is given by characteristic curves described by the Lagrange-Charpit equations:
\begin{equation}
    \dd{t}{\tau} = 1, \quad \dd{x}{\tau} = -kAh(\tau)^{k-1}, \quad \dd{h}{\tau} = u(x)
\end{equation}
The solution for curves starting at $t=0$ from $x=x_0\in[x_\up, L_x]$ with an initial surface height $h=0$ is given by:
\begin{equation}
    \label{eq:char_sol_1}
    t(\tau)=\tau, \quad x(\tau) = x_0 - A\frac{(r\tau)^k}{r}, \quad h(\tau) = r\tau,
\end{equation}
until they reach $x(\tau) = x_\up$ at $\tau=\tau_\up$. Afterwards, $h(\tau)$ remains constant since $u(x)=0$ for $x<x_\up$. From~\eqref{eq:char_sol_1}b, we find that
\begin{equation}
    \tau_\up = \frac{1}{r}\left[\frac{r(x_0-x_\up)}{A}\right]^{1/k}\quad\text{and}\quad h_\up=h(\tau_\up)=\left[\frac{r(x_0-x_\up)}{A}\right]^{1/k}.
\end{equation}
The subsequent characteristic solution for $\tau>\tau_k$ (equivalently $x<x_\up$) is given by:
\begin{equation}
    \label{eq:char_sol_2}
    t(\tau)=\tau, \quad x(\tau) = x_\up - kAh_\up^{k-1}(\tau-\tau_\up), \quad h(\tau) = h_\up.
\end{equation}
However, these curves intersect with the characteristic curves starting from $x_0<x_\up$, where $h(\tau)=0$ and $x(\tau)=x_0$ (see \cref{fig:shock_diagram}a). The intersection of two characteristic curves with different $h$ values results in the appearance of a shock wave with a discontinuous $h$ value (see \cref{fig:shock_diagram}c). Let us denote the location of this discontinuity at time $t$ as $s(t)$. The propagation of the shock is governed by the Rankine-Hugoniot condition, which in this case is written as:
\begin{equation}
    \label{shock_propagation}
    \dd{s}{t} = \frac{q(x=s,t)}{h(x=s,t)} = -Ah(x=s,t)^{k-1}
\end{equation}

The value $h(x,t)$ can be found by solving~\eqref{eq:char_sol_2}b, which can be rewritten as
\begin{equation}
    \label{eq:shock_front}
    x = x_\up - kAh(x,t)^{k-1}\left(t-\frac{h(x,t)}{r}\right).
\end{equation}
For $k>1$, this equation has two solutions for $h(x,t)$, but only the higher solution corresponds to the characteristic curve reaching the given point. The lower solution corresponds to the second time the characteristic curve would cross the shock wave, if it has not stopped at the shock wave before.

Solving~\eqref{eq:shock_front} for $h(x,t)$ allows us to find the propagation of the front from~\eqref{shock_propagation}. The solution, represented by the red line in \cref{fig:shock_diagram}, perfectly reproduces the numerical solution of the full Saint Venant equation~\eqref{eq:St_Venant}.

\section{Solution for the dry scenario}
\label{app:dry_scenario}

\subsection{Physical model}

We model the drying process using the benchmark scenario from~\cite{paper3}. We start from the same initial condition (steady state for a mean rainfall $\rho_0$), but we assume no precipitation $\rho=0$ after $t=0$.

\begin{figure}
    \centering
    \includegraphics{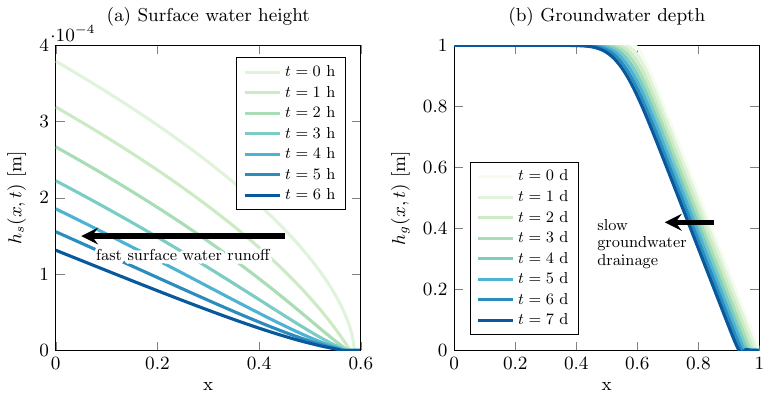}
    \caption{Solution for surface and subsurface flow during the dry scenario. Note the difference in timescale between overland and groundwater flow.}
    \label{fig:drying_solution}
\end{figure}

In the numerical results shown in \cref{fig:drying_solution}, we can observe that both the surface water and groundwater height decay over time. The decreasing groundwater height causes the seepage zone to shrink; however, this process is very slow and does not impact the river inflow until the seepage zone disappears completely. Therefore, we will not model this process. In this appendix, we derive an analytic expression for the overland component of the river inflow as a function of time, following the same argumentation as in Section 6.2.1 of our earlier work~\citep{paper3}.

We consider the dimensionless overland model from the aforementioned paper (eqns 6.4-6.5), but with $\rho=0$:
\begin{equation}
    \label{eq:dhdt_overland}
    \pt{h} - kh^{k-1}\px{h} - \Pe ^{-1}\ppx{h}= \rho = 0,
\end{equation}
with the initial condition:
\begin{equation}
    \label{eq:hs_initial}
    1 + h_0^k - \Pe^{-1}h_0'= \rho_0 (1 - x_0),
\end{equation}

Taking $\text{Pe}\rightarrow\infty$, equation~\eqref{eq:dhdt_overland} reduces to a first-order PDE:
\begin{equation}
    \label{eq:dhdt_overland_inf_Pe}
    \dt{h} - kh^{k-1}\dx{h} = 0,
\end{equation}
and the initial condition~\eqref{eq:hs_initial} becomes:
\begin{equation}
    \label{eq:hs_initial_inf_Pe}
    h_0^k = \rho_0 (a_0 - x_0),
\end{equation}
where $a_0=1-\frac{1}{\rho_0}$ is the size of the initial seepage zone.

We solve equation~\eqref{eq:dhdt_overland_inf_Pe} using the method of characteristics:
\begin{equation}
    \dd{t}{\tau} = 1 \quad\quad \dd{x}{\tau} = -k h^{k-1} \quad\quad \dd{h}{\tau} = 0
\end{equation}
with the solution:
\begin{equation}
    t(\tau)=\tau \quad\quad x(\tau) = x_0-k h_0^{k-1}\tau \quad\quad h(\tau) = h_0
\end{equation}

Now, by combining the above equations, we find the time $t^*$ at which surface water of a given height $h_0(x_0)$ reaches the river ($x=0$):
\begin{equation}
    0 = x_0 - k h_0(x_0)^{k-1} t^*
\end{equation}
which gives us:
\begin{equation}
    \label{eq:t_star_1}
    t^* = \frac{x_0}{k h_0(x_0)^{k-1}}
\end{equation}
From equation~\eqref{eq:hs_initial_inf_Pe}, we have:
\begin{equation}
    \label{eq:hs_initial_x1}
    x_0 =  a_0 - \frac{h_0^k}{\rho_0},
\end{equation}
By substituting~\eqref{eq:hs_initial_x1} into~\eqref{eq:t_star_1}, we get:
\begin{equation}
    t^*\left(h_0\right) = \frac{\rho_0a_0 - h_0^k}{k \rho_0h_0^{k-1}}
\end{equation}
or
\begin{equation}
    \label{eq:inverted_hydrograph}
    t^*\left(q^*\right) = \frac{\rho_0a_0 - q^*}{k \rho_0\left(q^*\right)^{1-\frac{1}{k}}}
\end{equation}
where $q^*=h_0^k$ is the dimensionless surface flow. This equation describes the time after which the overland flow drops to $q^*$. Its inversion would give us a hydrograph $q^*\left(t^*\right)$, but as in the case of the storm hydrograph, we cannot find an analytic expression for this inverse function. However, we can approximate its behaviour at early and late times.

For $t*\rightarrow 0$, we have $q^*\rightarrow a_0\rho_0$. By taking the leading-order approximation of~\eqref{eq:inverted_hydrograph} around $h^*=a_0\rho_0$, we get:
\begin{equation}
    t^*\left(q^*\right) = \frac{\rho_0a_0 - q^*}{k \rho_0\left(a_0\rho_0\right)^{1-\frac{1}{k}}},
\end{equation}
and its inverse function is:
\begin{equation}
    q^*\left(t^*\right) = \rho_0a_0 - k \rho_0\left(a_0\rho_0\right)^{1-\frac{1}{k}}t^* = \rho_0a_0\left(1 - \frac{k \rho_0}{\left(\rho_0a_0\right)^{1/k}}t^*\right).
    \label{eq:qs_early_time}
\end{equation}
Therefore $q-\rho_0a_0\propto t$ as $t\rightarrow 0$.

For $t*\rightarrow \infty$ we have $q^*\rightarrow 0$. By taking the leading order approximation of~\eqref{eq:inverted_hydrograph} around $h^*=0$, we get:
\begin{equation}
    t^*\left(q^*\right) = \frac{a_0}{k \left(q^*\right)^{1-\frac{1}{k}}}
\end{equation}
Its inverse function is:
\begin{equation}
    q^*\left(t^*\right) = \left(\frac{a_0}{k t^*}\right)^{\frac{k}{k-1}}
    \label{eq:qs_late_time}
\end{equation}
If $k=\frac{5}{3}$, then $q\propto t^{-5/2}$ as $t\rightarrow\infty$.

As shown in \cref{fig:dry_scenario}, both the approximated analytic solution~\eqref{eq:inverted_hydrograph} and its early- and late-time approximations~\eqref{eq:qs_early_time} and~\eqref{eq:qs_late_time} are consistent with the numerical results for the 1D model. Interestingly, the speed at which the seepage zone shrinks does not have any observable impact on the hydrograph (for a week-long drought simulation)---the hydrograph's shape can be fully explained by the dynamics of the surface water in the initial seepage zone alone.

\subsection{Grid and Grid-to-Grid models}

The drying process in the Grid model is described using a kinematic wave approximation~\eqref{eq:kinematic_wave_general} with a source term $u(x,t)=0$:
\begin{equation}
    \label{eq:KWA_dry}
    \pt{q(x,t)}-c\px{q(x,t)}=0,
\end{equation}
and with the initial condition:
\begin{equation}
    \label{eq:hs_initial_G2G}
    q_0(x)=r_0(L_x-x)
\end{equation}

This equation can be solved using the method of characteristics. The characteristic curves are governed by the following Lagrange-Charpit equations:
\begin{equation}
    \dd{t}{\tau} = 1 \quad\quad \dd{x}{\tau} = -c \quad\quad \dd{q}{\tau} = 0.
\end{equation}
Their solutions are given by:
\begin{equation}
    t(\tau)=\tau, \quad\quad x(\tau) = x_0-c\tau, \quad\quad q(\tau) = q_0(x_0)=r_0(L_x-x_0).
\end{equation}
From these equations, we can determine the overland flow $q(t)$ at the location of the river ($x=0$):
\begin{equation}
    q(t) = r_0(L_x-ct)
\end{equation}
Therefore, the overland flow $q(t)$ decreases proportionally with time from $r_0L_x$ to $0$ at $t=t_\mathrm{dry}=\frac{L_x}{c}$. After that, the overland flow remains constant.

In practice, there are two effects that are not included in this analysis:
\begin{enumerate}
    \item~The numerical diffusion related to the numerical scheme used to solve the kinematic wave approximation causes the flow not to reach zero at $t=\frac{L_x}{c}$ but rather to decay exponentially afterwards. The characteristic timescale of this decay decreases to $0$ as the number of mesh elements tends to infinity.
    \item~In the Grid-to-Grid model, even with no rainfall, there is a non-zero source term in \eqref{eq:KWA_dry} due to the return flow, \emph{i.e.} the transfer of groundwater flow to the surface. However, this additional overland flow decays over a very long timescale characterising the groundwater flow, and therefore should be considered as part of the base flow rather than the storm flow.
\end{enumerate}

\end{document}